\documentclass[12pt]{article}
\usepackage{epsfig,latexsym,amssymb}
\title{Circular geodesics in Extremal Reissner Nordstrom Spacetimes}
\author{Parthapratim Pradhan\footnote{pppradhan5@rediffmail.com}
\\{\it Department of Physics}\\{\it Vivekananda Satabarshiki Mahavidyalaya} \\
{\it Manikpara, Paschim Medinipur 721513, India.}\\ and \\
Parthasarathi Majumdar\footnote{parthasarathi.majumdar@saha.ac.in} \\
{\it Saha Institute of Nuclear Physics} \\{\it Kolkata 700 064, India.}}
\begin{document}

\maketitle

\begin{abstract}

Circular null geodesic orbits, in extremal Reissner-Nordstrom
spacetimes, are examined with regard to their stability, and compared with
similar orbits in the near-extremal situation. Extremization of the effective
potential for null circular orbits shows the existence of a stable circular geodesic 
in the extremal spacetime, precisely {\it on} the event horizon, which coincides with its
null geodesic generator. Such an orbit also emerges as a global minimum of the
effective potential for circular {\it timelike} orbits. This type of
geodesic is of course absent in the corresponding near-extremal spacetime, as
we show here, testifying to differences between the extremal limit of a
generic RN spacetime and the exactly extremal geometry.   
   
\end{abstract}

\section{Introduction}

A number of features of near-extremal black hole spacetimes indicate the
absence of a smooth limit to extremality \cite{rac}. One aspect related to
black hole thermodynamics is the definition of the so-called Entropy Function
\cite{wald} used widely nowadays to match the `macroscopic' entropy of a class
of extremal black holes emerging in the supergravity limit of string theories,
to the `microscopic' entropy obtained from counting of string states
\cite{sen}. While the state counting is strictly restricted to BPS states, the
use of the Entropy Function is stymied by the fact that its existence depends
on that of a {\it bifurcation two-sphere} (in four spacetime dimensions). This
bifurcation sphere only exists away from extremality, which forces one to
begin with a {\it near}-extremal situation, and then proceed eventually to the
extremal {\it limit}. 

However, as has been suspected earlier \cite{ddr} and
succinctly pointed out recently
\cite{carrand}, the existence of this limit cannot be taken for granted. In
other words {\it the extremal limit of a near-extremal spacetime is not
necessarily the extremal spacetime}. One manifestation of this concerns the
near horizon limit: Consider for instance the near-horizon geometry of a
non-extremal Reissner Nordstrom (RN) spacetime
\begin{eqnarray}
ds^2 = -\left[{(r-r_+)(r-r_-) \over r^2}\right] dt^2 +\left[{(r-r_+)(r-r_-)
    \over r^2}\right]^{-1} dr^2 + r^2 d\Omega^2 ~.
\label{rn}   
\end{eqnarray}
Near extremality, defining $\delta \equiv (r_+-r_-)/r_0 << 1~,~\epsilon \equiv
(r-r_0)/r_0 <<1$ where $r_0$ is the radius of the horizon in the extremal case,
this metric reduces, close to the event horizon, to
\begin{eqnarray}
ds^2 = -{\epsilon (\epsilon + \delta) \over (1+\epsilon)^2} dt^2 + {(1+\epsilon)^2
\over \epsilon(\epsilon + \delta)} dr^2 + r_0^2 (1+\epsilon)^2 d\Omega^2
~\label{near}
\end{eqnarray} 
which leads to two distinct outcomes depending on the order in which the
limits $\delta \rightarrow 0$ and $\epsilon \rightarrow 0$ are taken. If the
extremal limit $\delta \rightarrow 0$ is taken first and then the near-horizon
limit is taken, the {\it local} geometry is that of an $AdS_2 \times S_2$ 
\begin{eqnarray}
ds^2 \simeq -\epsilon^2 dt^2 + {r_0^2 \over \epsilon^2} d\epsilon^2 + r_0^2
d\Omega^2~. 
\label{ads2}
\end{eqnarray}
On the other hand, if the near horizon limit is taken before the extremal limit, one gets
\begin{eqnarray}
ds^2 \simeq - \epsilon \delta dt^2 + {r_0^2 \over \epsilon \delta } d\epsilon^2
+ r_0^2 d\Omega^2 ~, \label{nearh}
\end{eqnarray}
which certainly does not correspond to an $AdS_2 \times S_2$; in fact the
extremal limit is now singular.

Indeed, it is known \cite{rac} that extremal spacetimes do
not have any trapped surface inside the event horizon (itself usually a
marginal outer trapped surface). This makes use of the fact, first pointed out
in \cite{bek}, that the proper distance between the event and Cauchy horizons
in the extremal geometry (in the RN case, for instance) is
actually infinite, even though the coordinate distance vanishes. This is
certainly not the situation in the non-extremal situation where the proper
distance between the inner and outer horizons is finite, as is the coordinate
distance. The continued use of the extremal limit of a generic spacetime {\it
  as} the extremal geometry may not thus be above suspicion.

One aspect that has not been studied in detail is the behaviour of geodesics,
in the exterior of such spacetimes, both in the
near-extremal and extremal cases. The interest here is in what happens to the
geodesics near the horizon. There is a class of timelike geodesics in spherically symmetric
spacetimes like Schwarzschild and RN, which encircle the
event horizon in a stable orbit (the so-called ISCOs). There is however no
stable null geodesic orbit for any generic spacetime. Since our interest is
in the behaviour of circular null orbits near the event
horizon of an extremal RN black hole, the use of the Kruskal-Szekeres extension of such spacetimes becomes
crucial. Already at this level, there seems to be a discontinuity in the
extremal limit of the Kruskal-Szekeres extension of a generic RN spacetime, vis-a-vis
the extension of the extremal RN spacetime. It further ensues that circular
null geodesics near
the event horizon of an extremal RN spacetime exhibit a behaviour quite
different from that in near extremal situations. In
what follows, we consider the behaviour of such orbits in some detail for
near-extremal and extremal RN spacetimes. 

A further motivation for the work comes from Hawking radiation, which is known
to be absent for the extremal spacetime, as the surface gravity on the event
horizon which measures the equilibrium temperature for the thermal
distribution of the radiation vanishes in this case. Once again, this thermal
state cannot be achieved in a continuous manner from a radiant black hole (however weakly)
without violating energy conditions \cite{isr}. The behaviour of circular orbits is
relevant to this in order to ascertain what really happens at
extremality. This is an important issue for rotating black holes for which
black hole radiance also includes {\it superradiance} in addition to Hawking
radiation. While we do not consider rotating black holes in this paper, these
issues serve as motivation for the present work. 

The plan of the paper is as follows: in section 2 we exhibit the
Kruskal-Szekeres extension for non-extremal RN spacetime. We show
explicitly how the extremal limit of this extension is singular, implying that
extremal RN geometry has to be thus extended directly (e.g., \'a la Carter
\cite{carter1}), instead of by a limiting
procedure on the extension of the non-extremal geometry.  In section 3 we
calculate the radial location of a possible circular orbit for extremal RN spacetime,
and compare this with the radial location
of circular orbits in {\it near}-extremal RN black holes. We point out that there is a
{\it stable circular null orbit on the event horizon} in the extremal
spacetime, which disappears in the near extremal
geometry. This is a key result of our analysis. The concluding section (4)
includes a discussion of our results in the light of trapped surfaces and also
presents our future outlook.

\section{Kruskal-Szekeres extension of RN spacetime}

The Kruskal-extended RN spacetime was first worked out by Carter
\cite{carter1} who gave the extended geometry only for the extremal case. Here
we consider first the extension of the {\it non}-extremal or generic RN
spacetime. While the procedure is now standard textbook material, we have not
been able to locate in the literature an adequate discussion of the pathologies in this
extension, that manifest in the extremal limit $r_+ \rightarrow r_-$. The
singular behaviour discerned here bears an unmistakable stamp of the
subtleties of the extremal limit. In other words, we show that the {\it
  extremal limit} of the Kruskal extension is {\it not} the same as the
Kruskal-extended extremal RN spacetime found in \cite{carter1}.
 
\subsection{Non-extremal case}

We begin by defining `tortoise' coordinates, and then use that to derive the
Kruskal extension. The tortoise coordinate $r^\ast$ is given by
\begin{equation}
dr^\ast=\frac{r^2dr}{\Delta}=\frac{r^2dr}{(r-r_{-})(r-r_{+})}~.  \label{4.1}
\end{equation}
Integrating this equation, we obtain
\begin{equation}
r^\ast=r+\frac{r_{+}^2}{(r_{+}-r_{-})}\ln|r-r_{+}|
          -\frac{r_{-}^2}{(r_{+}-r_{-})}\ln|r-r_{-}|+c\label{4.3}
\end{equation}
where as usual $r_{\pm}=M\pm\sqrt{M^2-Q^2}$ and $c$ is an integration constant. The
outer horizon $r_{+}$ is an event horizon and the inner horizon $r_{-}$ is a Cauchy horizon.
Now, near the event horizon $r=r_{+}$, the tortoise coordinate is given by
\begin{equation}
r^\ast \approx \frac{r_{+}^2}{(r_{+}-r_{-})}\ln|r-r_{+}|\label{4.4}
\end{equation}
Here $r^\ast$  has logarithmic dependence on $r-r_+$ and is singular at
$r=r_{+}$. Introducing the radial null coordinates $u$ and $v$, given by
$u=t-r^\ast$, $v=t+r^\ast$, we observe that the surface $r=r_{+}$ appears at
$v-u=-\infty$. This implies that we need another transformation, which is
the Kruskal-Szekeres extension. The null coordinates 
\begin{eqnarray}
du=dt-dr^\ast &=& dt-\frac{r^2dr}{\Delta} \\
dv=dt+dr^\ast &=& dt+\frac{r^2dr}{\Delta} ~, \label{4.5}
\end{eqnarray}
since $dr^{\ast}=\frac{r^2dr}{\Delta}$, where $u=const.$ are outgoing radial
null geodesics and  $v=const.$ ingoing radial null geodesics respectively.
The RN metric now assumes the form
\begin{eqnarray}
ds^2=-\frac{\Delta}{r^2}(dt^2-{dr^\ast}^2)+r^2d\Omega^2 \nonumber\\
     = -\frac{\Delta}{r^2}dudv +r^2d\Omega^2\label{4.5.1}
\end{eqnarray}
where $dudv=dt^2-dr^{\ast 2}$.

The Kruskal-Szekeres frame is now defined by the transformations 
\begin{eqnarray}
U^{+} &=& -\exp{(-\kappa_{+} u)} \\ 
V^{+} &=& \exp{(\kappa_{+} v)} \label{4.6}
\end{eqnarray}
where $\kappa_{\pm}=\frac{r_{\pm}-r_{\mp}}{2r_{\pm}^2}$ is the surface gravity of the null hypersurfaces.

Therefore the metric, near $r=r_{+}$, becomes
\begin{equation}
ds^2=-\frac{r_{+}r_{-}}{\kappa_{+}^2}\frac{\exp{(-2\kappa_{+}r})}{r^2}
      (\frac{r_{-}}{r-r_{-}})^{\kappa_{+}/\kappa_{-}-1} dU^{+} dV^{+}+r_{+}^2d\Omega^2 \label{4.7}
\end{equation}
where
\begin{equation}
U^+ V^+=-\exp{(2\kappa_{+} r)}(\frac{r-r_{+}}{r_{+}})(\frac{r-r_{-}}{r_{-}})^{\frac{\kappa_{+}}{\kappa_{-}}} \label{4.8}
\end{equation}
This implies that, when in terms of the Kruskal coordinates $(U^{+},
V^{+})$, the metric is well behaved at the event horizon $(r=r_{+})$ but is
singular at the inner (Cauchy) horizon $(r=r_{-})$.  From this we can conclude
that these coordinates $(U^{+}, V^{+})$ are valid for the region $(r_{-}< r<
r_{+})$ which is different from the original patch covering the region
$r>r_{+}$. These coordinates do not cover $r\leq r_{-}$ because of
the singularity at $r=r_{-}$, so another new coordinate patch is required to
cover this region. In this region $g_{tt}>0$ and $g_{rr}<0$ such that $t$ is
spacelike and $r$ is timelike. Note that in the above metric if we take the extremal limit $ r_- \rightarrow r_+$
then the metric diverges, proving that the extremal limit is not continuous
insofar as this extension is concerned. This follows from the fact that the
coordinate chart $U^+,V^+$ considered here does not extend to the Cauchy horizon. 

Now consider what happens for the Cauchy horizon $r=r_{-}$. Near the Cauchy
horizon $r=r_{-}$, the tortoise coordinates is given by
\begin{equation}
r^\ast\approx\frac{r_{-}^2}{(r_{+}-r_{-})}\ln|r-r_{-}| \label{4.9}
\end{equation}
Here $r^\ast$  has a logarithmic singularity at $r=r_{-}$. The radial null
coordinates $u$ and $v$ are $u=t-r^\ast$, $v=t+r^\ast$, then the surface
$r=r_{-}$ appears at $v-u=+\infty$. Therefore the Kruskal-Szekeres transformations are
\begin{eqnarray}
U^{-} &=& -\exp{(-\kappa_{-} u)} \\
V^{-} &=& \exp{(\kappa_{-} v)} \label{4.10}
\end{eqnarray}
where $\kappa_{-}$ has been previously defined. Therefore near $r=r_{-}$, the
metric becomes
\begin{equation}
 ds^2=-\frac{r_{+}r_{-}}{\kappa_{-}^2}\frac{\exp{(-2\kappa_{-}r})}{r^2}
      (\frac{r_{+}}{r-r_{+}})^{\kappa_{-}/\kappa_{+}-1} dU^{-} dV^{-}+r_{-}^2d\Omega^2 \label{4.11}
\end{equation}
where
\begin{equation}
 U^- V^-=-\exp{(2\kappa_{-} r)}(\frac{r-r_{-}}{r_{-}})(\frac{r-r_{+}}{r_{+}})^{\frac{\kappa_{-}}{\kappa_{+}}} \label{4.12}
\end{equation}
This implies that, when expressed in terms of Kruskal coordinates $(U^{-},
V^{-})$, the metric is regular at the Cauchy (inner) horizon $(r=r_{-})$
but singular at the event horizon $(r=r_{+})$. From this we can conclude that
these coordinates $(U^{-}, V^{-})$ are valid for the region $(0< r< r_{-})$.
Once again, the extended metric is singular in the extremal limit, thwarting
any attempt to derive the Kruskal extension of the extremal spacetime by a
straightforward limiting procedure on the non-extremal extended geometry.

\subsection{Extremal case}

Now we want to see what happens for the extremal case. The tortoise coordinate is given by
\begin{equation}
r^\ast=\int\frac{r^2dr}{(r-M)^2}=r+2M[\ln|r-M|-\frac{M}{2(r-M)}] ~. \label{4.13}
\end{equation}
Near the horizon this becomes
\begin{equation}
r^\ast\approx\frac{M^2}{(r-M)} \label{4.14}
\end{equation}
Here $r^\ast$  has no logarithmic dependence, but an extra pole : $M^2/(r-M)$;
it is singular at $r=M$. Introducing the double null coordinates $u$ and $v$
as $u=t-r^\ast$, $v=t+r^\ast$, the surface $r=r_{+}$ appears at $v-u=\infty$,
hence these coordinates are inappropriate there. We need another
transformation as in the previous generic cases.

The metric in terms of double null coordinates $u$ and $v$ is given by
\begin{equation}
ds^2=-(1-M/r)^2dudv +r^2(u,v)d\Omega^2 \label{4.15}
\end{equation}
Now from eqn. (\ref{4.13})
\begin{equation}
\exp{(\alpha r^*)}=\exp{(\alpha r)}(r-M)^2\exp{(-M/(r-M))}~,~\alpha=1/M \label{4.16}
\end{equation}
The maximally extended spacetime can be obtained by substituting
\begin{eqnarray}
\tan U &=&-\exp{(-\alpha u)}=-\exp{(-\alpha t)}[\exp{(\alpha
  r)}(r-M)^2\exp(-M/(r-M))] \nonumber \\
\tan V &=&+\exp{(+\alpha v)}=\exp{(\alpha t)}[\exp{(\alpha r)}(r-M)^2\exp(-M/(r-M))] \label{4.17}
\end{eqnarray}

Therefore the complete extremal RN metric in Kruskal-Szekeres coordinate system
is given by (after substituting the value of $\alpha=1/M$)
\begin{eqnarray}
ds^2=-4M^2(1-\frac{M}{r})^2 \csc 2U \csc 2V dU dV+r^2d\Omega^2\nonumber\\
\tan U \tan V=-\exp(2r/M)(r-M)^4\exp(-2M/(r-M))\label{4.18}
\end{eqnarray}
This is the result of Carter \cite{carter1}.

\section{Circular Orbits in RN spacetime}

\subsection{Extremal spacetime}

As is well-known, The RN spacetime has a timelike Killing vector $\xi \equiv \partial_t$ whose
projection along the 4-velocity ${\bf u}$ (${\bf u}^2=-1$ for timelike and
${\bf u}^2=0$ for null) of geodesics: $\xi \cdot
{\bf u} = -E$, is conserved along such geodesics. Recall that the timelike Killing
vector becomes null on the event horizon. There is also the `angular
momentum' $L \equiv \zeta \cdot {\bf u}$ (where $\zeta
\equiv \partial_{\phi}$) which is similarly conserved. It is straightforward to show that
(see, e.g., \cite{scbh}) for circular null orbits ($r(U,V)=R$), $E$ obeys the equation
\begin{eqnarray}
E^2 = {\cal U}_{eff}(R) = \left({L^2 \over R^2} \right)~\left(1-{M \over
    R} \right)^2~, \label{ueff}
\end{eqnarray}   
where ${\cal V}_{eff}(R)$ is called the effective potential. For timelike
orbits, the corresponding effective potential is given by 
\begin{eqnarray}
E^2 = {\cal V}_{eff}(R) = \left(1 + {L^2 \over R^2} \right)~\left(1-{M \over
    R} \right)^2~. \label{veff}
\end{eqnarray} 
If this effective
potential has an absolute minimum, this is identified with the radial location
of a {\it stable} circular orbit. Alternatively, other extrema are identified as radial locations of
unstable circular orbits.

It is convenient to define dimensionless quantities ${\bf r} \equiv R/M~,~\ell \equiv L/M$ and $q \equiv
Q/M$. Thus, the extremal case corresponds to $q=1$. 

\subsubsection{Null orbits}

In this notation, the effective potential becomes 
\begin{eqnarray}
{\cal U}_{eff}({\bf r}) = {\ell^2 \over {\bf r}^2} ~ \left(1 - {1
    \over {\bf r}} \right)^2 ~. \label{uefft}
\end{eqnarray}
Setting ${\cal U}_{eff}'({\bf r}) = 0$ one obtains 
\begin{eqnarray}
{2\ell^2 \over {\bf r}^3} \left(1 - {1 \over {\bf r}} \right) \left({2 \over
    {\bf r}} -1 \right) = 0~. \label{solnn}
\end{eqnarray}
which has the solutions as circular orbits at ${\bf r} = 1,2$. 

It is easy to check that
${\cal U}_{eff}''(1) > 0$ which corresponds to a global minimum of ${\cal
U}_{eff}$. This can be taken to correspond to a stable circular orbit for a
photon with $E=0$. On the other hand, ${\cal U}_{eff}''(2) < 0$, implying that there is
no other stable circular photon orbit.   

Thus, there is a null stable circular orbit {\it on} the event horizon of an extremal RN
spacetime, a feature which appears to be novel. We have not seen any
discussion of this type of orbit in extremal RN spacetime anywhere in the
literature. This circular geodesic can be seen to coincide
with the null geodesic generator of the event horizon. The real novelty here
is that it appears as a global minimum of the effective
potential. While a null geodesic generator of the event horizon is not thought
of as a stable circular orbit, since there is no such orbit in generic black
hole spacetimes (derived as an absolute minimum of the effective potential),
we would like to point out that the extremal RN is special in
this sense. However, we expect similar orbits to exist in other extremal
spacetimes as well. Before considering whether such an orbit exists in the
{\it near} extremal spacetime, we record an interesting finding for timelike orbits.

\subsubsection{Timelike orbits}

The effective potential in this case is given by
\begin{eqnarray}
{\cal V}_{eff}({\bf r}) = \left( 1 + {\ell^2 \over {\bf r}^2}\right) ~ \left(1 - {1
    \over {\bf r}} \right)^2 ~. \label{vefft}
\end{eqnarray}
Setting ${\cal V}_{eff}'({\bf r}) = 0$ one obtains 
\begin{eqnarray}
{\bf r}^2 ({\bf r} -1) - \ell^2 ({\bf r} -2) ({\bf r} -1) = 0 ~, \label{extr}
\end{eqnarray}
\begin{eqnarray}
{2\ell^2 \over {\bf r}^3} \left(1 - {1 \over {\bf r}} \right) \left({2 \over
    {\bf r}} -1 \right) = 0~. \label{solnt}
\end{eqnarray}
The solutions are given by
\begin{eqnarray}
{\bf r}_0 &=& 1 \nonumber \\
{\bf r}_{\pm} &=& {\ell^2 \over 2} \left[ 1 \pm \left( 1- {8 \over \ell^2}
  \right)^{1/2} \right] ~\label{solts}
\end{eqnarray}

It is easy to check that the circular orbit with ${\bf r} ={\bf r}_0 = 1$ {\it indeed
corresponds to the likely position of a stable circular orbit},
being a stable global minimum of ${\cal V}_{eff}$ given in
eq. (\ref{vefft}), since ${\cal V}_{eff}''(1) > 0$. This orbit is located
exactly as the null orbit found above at the event horizon, and must be the
same orbit, even though it shows up as a global minimum of the potential for
timelike orbits. The likely reason is that the Killing vector field which is timelike
everywhere in the exterior RN spacetime, turns null on the event horizon.  

Of the two other orbits at ${\bf r}={\bf r}_{\pm}$, clearly one must choose $\ell^2
\geq 8$ for them to be real; we choose $\ell^2=9$ for simplicity and get ${\bf
  r}_+=6~,~{\bf r}_-=3$. One obtains ${\cal V}_{eff}''(6) > 0$, showing that
the orbit with radius ${\bf r}_+$ is a stable {\it local} minimum, possibly
the ISCO, while ${\cal V}_{eff}''(3) < 0$, so that the orbit with radius ${\bf r}_-$ is unstable. 

The issue is now: what happens to the orbit on the event horizon in the
near extremal case, when we move infinitesimally away from
extremality in the black hole parameter space. To this we turn in the next
subsection.  

\subsection{Near extremal RN spacetime}

\subsubsection{Null Orbits}

The effective potential for null circular geodesics in generic non-extremal RN spacetime is determined
exactly as in the last subsection. In terms of the dimensionless quantities
introduced in that section, this effective potential can be expressed as
\begin{eqnarray}
{\cal U}_{eff}({\bf r}) = {\ell^2 \over {\bf r}^2} ~ \Delta({\bf
  r}) \label{nxrn}
\end{eqnarray}     
where, 
\begin{eqnarray}
\Delta({\bf r}) \equiv 1 - \frac{2}{\bf r} + {q^2 \over {\bf r}^2} ~.
\label{rndel}
\end{eqnarray}
We solve the extremization equation ${\cal U}_{eff}'({\bf r}) =0$
perturbatively around the extremal solutions satisfying eq. (\ref{solnn}),
corresponding to the {\it near extremal} geometry. To this end, we define
$\chi \equiv 1-q >0 ~,~\rho \equiv {\bf r} -1$, where, recall
that for the extremal situation $q=1$, the stable circular orbit is at ${\bf r} = {\bf
r}_0=1$. In our chosen units, the event horizon is located at $r_+ = 1 + (2\chi)^{1/2}
+ O(\chi^{3/2})$ while the Cauchy horizon is at $r_- = 1 - (2\chi)^{1/2}
+ O(\chi^{3/2})$. The extremization equation equation is to be solved perturbatively around
the extremal solution to yield $\rho$ to {\it linear} order in the
perturbation $\chi$. Likewise, terms of $O(\rho^2)$ or higher 
are ignored. In terms of the linear perturbation $\chi$, the effective potential
for null circular orbits is given by 
\begin{eqnarray}
{\cal U}_{eff}({\bf r}) = {\ell^2 \over {\bf r}^2 } \left [ \left(1- {1 \over
      {\bf r}} \right)^2 - {2\chi \over {\bf r}^2} \right]    ~. \label{ueffn}
\end{eqnarray}
The extremization equation ${\cal U}_{eff}'({\bf r})=0$ yields the quadratic
equation 
\begin{eqnarray}
{\bf r}^2 -3{\bf r} + 2(2-\chi) =0~. \label{extrnn}
\end{eqnarray} 
leading to circular orbits with radii ${\bf r}=1-4\chi~,~2 +
4\chi$. It is easy to show that ${\cal U}_{eff}''({\bf r}) < 0$ for both
these radii, demonstrating that the near extremal spacetime does not admit any
stable null circular orbit at all. This is merely a rephrasing of what is
well-known in the literature \cite{scbh} : the generic RN spacetime does not
admit a stable null circular orbit, just like the Schwarzschild spacetime. The
excercise however underlines the key point we wish to make in this paper :
{\it the stable circular null geodesic on the event horizon present in the
extremal case as the global minimum of the effective potential, is absent in
the near extremal case}. This adds to the list of disparities between the
extremal limit of the generic RN spacetime and the actual extremal spacetime.   

For completeness, we include a treatment for timelike orbits in the near
extremal geometry.

\subsubsection{Timelike orbits}

In terms of the dimensionless quantities
introduced in that section, this effective potential can be expressed as
\begin{eqnarray}
{\cal V}_{eff}({\bf r}) = \left(1 + {\ell^2 \over {\bf r}^2} \right) ~ \Delta({\bf
  r}) \label{nxrn}
\end{eqnarray}     
where, $\Delta({\bf r})$ is given above in (\ref{rndel}). The extremization
condition ${\cal V}_{eff}'({\bf r}) =0$ leads to the cubic equation
\begin{eqnarray}
{\bf r}^3 -(q^2 + \ell^2) {\bf r}^2 + 3\ell^2 {\bf r} -2q^2 \ell^2 = 0 ~.
\label{minm}
\end{eqnarray}

It is sufficient for us to solve eq. (\ref{minm}) in the {\it near extremal}
approximation, to compare the results with those in the extremal situation.
One gets, to linear order in the perturbation
\begin{eqnarray}
\rho = -2\chi \left( {1 +2\ell^2} \over {1 + \ell^2} \right) + O(\chi^2)
~, \label{unstab}
\end{eqnarray}
leading to a circular orbit with radius
\begin{eqnarray}
{\bf r}_0^{nex} \simeq 1 - 2\chi \left( {1 + 2\ell^2} \over {\ell^2 + 1} \right) <
r_+ ~! ~\label{non}
\end{eqnarray}
Thus, the stable circular orbit at ${\bf r}_0=1$ found for the extremal spacetime has
no analogue in the near-extremal situation. 

To find a stable circular orbit in this case, we proceed as before. Perturbing around the
unstable orbit at ${\bf r}_- = (1/2)\ell^2 [ 1 - (1-8/\ell^2)^{1/2} ]$ we obtain
the perturbation 
\begin{eqnarray}
\rho_- = -2 \chi \left[ {\ell^2 {\bf r}_- \over {(\ell^2 -2) {\bf r}_- -3
\ell^2 }} \right]~. \label{rho-}
\end{eqnarray}
It is obvious that with $\ell^2 > 8$ one gets $\rho_- < 0$, this orbit does indeed
correspond to an unstable circular orbit in the near extremal spacetime. In
fact, for $\ell^2=9~,~ \rho_- = 9\chi$.  On
the other hand, perturbation around the stable circular orbit at 
${\bf r}_+ = (1/2)\ell^2 [ 1 + (1-8/\ell^2)^{1/2} ]$ in the extremal case,
leads to a perturbed orbit with 
\begin{eqnarray}
\rho_+ = -2 \chi \left[ {\ell^2 {\bf r}_+ \over {(\ell^2 -2) {\bf r}_+ -3
\ell^2 }} \right]~. \label{rho+}
\end{eqnarray}
It is not difficult to see that for $\ell^2 > 8$ we have $\rho_+ > 0$. In fact, for
$\ell^2=9$, $\rho_+= -(36/5) \chi$. This does indeed correspond to a stable circular
orbit in the near extremal spacetime, indeed it is the ISCO in this case.  
 
\section{Discussion}

The class of stable circular orbits on the event horizon of the extremal RN
spacetime, discerned above, coincides with the null geodesic generator of the horizon. However,
what is interesting is that this happens only in the extremal spacetime. For the
near extremal geometry, the effective potential has {\it no global minimum}
corresponding to such a geodesic. Indeed, neither for the Schwarzschild nor
for any other generic spherically symmetric spacetime is the null geodesic
generator on the event horizon represented by a {\it stable} circular orbit
{\it on} the horizon, characterized by an absolute minimum of the effective
potential for circular orbits. The fact that such a class of orbits is absent
even in the near extremal geometry is another illustration of the subtlety
associated with the extremal {\it limit}. This can be seen to have arisen from
the absence of outer trapped surfaces within the horizon in the extremal
geometry in contrast to a more generic situation.
 
\subsection{Absence of trapped surfaces in extremal RN spacetime}

In a general spacetime $(M,g_{\mu\nu})$  with the metric $g_{\mu \nu}$ having
signature $(-+++)$, one defines two future directed null vectors $l^{\mu}$ and
$n^{\mu}$ whose expansion scalars are given by
\begin{eqnarray}
\theta_{(l)}=q^{\mu\nu}\nabla_{\mu}l_{\nu} \,\,\,
\theta_{(n)}= q^{\mu\nu}\nabla_{\mu}n_{\nu} ~. \label{expa}
\end{eqnarray}
where $q_{\mu\nu}=g_{\mu\nu}+l_{\mu}n_{\nu}+n_{\mu}l_{\nu}$ is the metric
induced by $g_{\mu \nu}$ on the two dimensional spacelike surface formed by
spatial foliation of the null hypersurface generated by $l^{\mu}$ and $n^{\mu}$.

Then (i) a two dimensional spacelike surface S is said to be a {\it trapped} surface if both $\theta_{(l)}<0$ and
$\theta_{(n)}<0$; (ii) S is to be {\it marginally trapped} surface if one of two null
expansions vanish i.e. $\theta_{(l)}=0$ or $\theta_{(n)}=0$. The null vectors
for non-extremal RN black hole are given by
\begin{eqnarray}
l^{\mu}=\frac{1}{\Delta}(r^2,-\Delta,0,0) \,\,\, n^{\mu}=\frac{1}{2r^2}(r^2,\Delta,0,0)\label{7.2}
\end{eqnarray}
\begin{eqnarray}
l_{\mu}=\frac{1}{\Delta}(-\Delta,-r^2,0,0) \,\,\, n_{\mu}=\frac{1}{2r^2}(-\Delta,r^2,0,0)\label{7.3}
\end{eqnarray}
where $\Delta=(r-r_{+})(r-r_{-})$ and $r_{\pm}=M\pm\sqrt{M^2-Q^2}$.
The null vectors satisfy the following conditions :
\begin{equation}
l^{\mu}n_{\mu}=-1 \,\, l^{\mu}l_{\mu}=0 \,\, n^{\mu}n_{\mu}=0\label{7.4}
\end{equation}

Using (\ref{expa}), one obtains
\begin{equation}
\theta_{(l)}=-\frac{2}{r} \,\, \theta_{(n)}=\frac{(r-r_{+})(r-r_{-})}{r^3}\label{7.5}
\end{equation}
In the region $(r_{-}< r< r_{+})$,  $\theta_{(l)}<0$ and
$\theta_{(n)}<0$. This implies that trapped surfaces exist for non extreme RN
black hole in this region. In contrast, for the extreme RN black hole
\begin{equation}
\theta_{(l)}=-\frac{2}{r} \,\, \theta_{(n)}=\frac{(r-M)^2}{r^3}\label{7.6}
\end{equation}
Here inside or outside extremal horizon $r<M$ or $r>M$, $\theta_{(l)}<0$ and
$\theta_{(n)}>0$. 
This implies that there are no trapped surfaces for extremal RN black hole
beyond the event horizon .

\subsection{Outlook}

Our analysis reinforces earlier assertions in the literature \cite{ddr},
\cite{carrand} that the extremal limit of a generic charged black hole is {\it
  not necessarily} the extremal black hole spacetime. However, recent work
\cite{ayan} has conclusively shown that extremal black holes can be modeled
by {\it isolated horizons} on par with generic non-extremal black holes. One
expects this to lead to
a well-defined microcanonical entropy obtained for extremal macroscopic black holes
as an infinite series in horizon area, with the leading
Bekenstein-Hawking area term receiving precise subleading logarithmic and power law
corrections \cite{km2}, for spherically symmetric horizons, just like more
generic black holes. Note that this is
genuine {\it gravitational} entropy of extremal black holes, and has little to
do with entanglement or such non-gravitational phenomena. This
understanding of extremal black hole entropy, based on Loop Quantum Gravity,
ought to find extensive applications in superstring theoretic black holes in four
dimensional spacetime \cite{ayan2}. The object now is to discern whether this
approach works for rotating black holes
with a similar degree of precision. Apart from questions pertaining to geodesics
and circular orbits in extremal Kerr and Kerr-Newman spacetimes, there is the issue of
stability of such spacetimes with respect to {\it superradiance}. Isolated
horizons are not expected to be very useful in this respect, since the
properties of the ergosphere play a crucial role for superradiance. The aim in
future then ought to be a study of perturbations of extremal Kerr black holes
with regard to superradiance.      

\vglue .5cm

\noindent {\bf Acknowledgment :} We thank R. Basu, A. Chatterjee and A. Ghosh
for helpful discussions.

\end{document}